# Hidden Quantum State and Signature of Mott Transition in Two-dimensional 1T-TaS$_2$

Vivek Kumar[*,1], Birender Singh[*,2] and Pradeep Kumar[*,3]

[*]School of Physical Sciences, Indian Institute of Technology Mandi, Mandi-175005, India

**Abstract**

Here we report a comprehensive inelastic light scattering studies on 1T-TaS$_2$ with different thickness. This compound is well known for its rich charge density wave phases. Along with that it has been one of the promising candidates for a quantum spin liquid state as the spins reside on a triangular lattice and it does not show any signature of magnetic ordering. We performed a thickness dependent Raman measurement in a regime of completely commensurate charge density wave (C-CDW) to a nearly commensurate charge density wave (NC-CDW) with varying temperature (4K-330K) and polarization direction of the incident light. We observed the signature of CDW transition and in addition to that we have also found the signature of a well sought hidden quantum CDW state at low temperature around T$_H$ ~ 80K. The emergence of CDW, both normal and hidden one, is marked by the emergence of new phonon modes and distinct renormalized phonon self-energy parameters for the most prominent modes. A transition from metallic to the Mott insulating state is gauged via the Raman response using low frequency slope $\left(\lim_{\omega \to 0} \partial \chi'' / \partial \omega \right)$, reflected in the renormalized slope below T$_{CDW}$ and T$_H$.

[1]vivekvke@gmail.com
[3]pkumar@iitmandi.ac.in
[2]Current affiliation: Department of Physics, Boston College, Chestnut Hill, MA, USA.

**Introduction**

Quasi-two-dimensional (2D) transition metal dichalcogenides (TMDs) have been a center of interest in scientific community for past several decades due to their rich properties. Group V TMDs (V, Nb, Ta and Tb), specifically have been known to feature charge density waves (CDWs) and superconductivity. CDWs and superconductivity are two such electronic ground states that often compete and/or coexist in materials such as Cuprates and TMDs. Such an intriguing interplay has been vastly studied theoretically as well as experimentally, but still it has perplexed scientists in the complete understanding of such a complex phenomenon [1-3]. The CDW transitions in these materials occur at relatively higher temperatures which is a favorable feature to investigate hysteresis effects and is a defining aspect for potential applications such as data storage [4], information processing [5], or voltage-controlled oscillators [6].

Generally, the factors that dictate CDW phase transitions in two-dimensional and three-dimensional systems are Fermi surface nesting [7,8], electron-phonon coupling [9,10], and excitonic insulator [11,12]. In 1930, Rudolf Peierls first proposed a new kind of phase transition wherein the one-dimensional metals turn insulator resulting in a gap opening at the Fermi-level ($\pm K_F$) owing to the inbuilt lattice instability at low temperatures [13]. Later in 1954, Herbert Fröhlich proposed a detailed microscopic theory of 1-D superconductivity which predicted CDW formation and resulting in collective charge transport below transition temperatures [14] . In CDW state the periodic lattice instability is accompanied by fluctuations in the electron density which can be described phenomenologically as $\rho(\vec{r}) = \rho_o + \rho_A \cos(2\vec{K}_F \cdot \vec{r} + \theta)$ below the transition; here $\rho_o$ is undistorted lattice electron density, $\rho_A$ is the amplitude of the modulations and $\theta$ is the phase of the CDW that determines its position relative to the underlying lattice. It can be characterized by an order parameter $\xi = \Delta e^{i\theta}$, the magnitude '$\Delta$' dictates the electronic energy gap and atomic displacement. The modulation in $\theta$ and $\Delta$ give rise to collective excitations as phasons and amplitudons, respectively; and have been observed in Raman spectroscopic measurements [15]. In commensurate phase (C-CDW) the altered periodic phase lattice constant is an integer multiple of the undistorted one, where $\cos\theta = \pm 1$ and also known as pinned phase. Brillouin zone folding effects can be seen in this phase as emergence of new phonon modes at $\Gamma$ point. On the other hand, if it is not an integer multiple then it is known as incommensurate phase (IC-CDW). Here, charge density wave is independent of the choice of phase or position thus could move and hence the Peierls gap in the k-space which induces an electric current which is proportional to $\partial\theta/\partial t$. This phase lacks translational symmetry and the vibrational modes are no longer plane-waves, hence the corresponding relaxation of the phonon momentum conservation mimics to that of amorphous materials with broad Raman peaks [16,17]. In a nearly or partially commensurate phase (NC-CDW), the IC-CDW and C-CDW coexist [18,19].

Among the TMDs 1T-TaS$_2$ is known to be a correlation driven Mott insulator. Owing to an imperfect nesting in 2D the CDW gaps out only certain part of the Fermi surface which leaves behind a metallic state that often becomes superconducting [20]. In a standard description of Mott insulator where an anti-ferromagnetic (AFM) ground state is formed due to the exchange coupling. As far as the magnetic properties of 1T-TaS$_2$ is concerned there is no such ordering of spins has been reported. Magnetic susceptibility drops at CDW transitions with decreasing temperatures but remains nearly constant below 200K which is unlike Curie Weiss behavior [20-22]. It has been advocated that this material shows a quantum spin liquid state, either a fully gapped Z2 spin liquid or a Dirac spin-liquid [21]. Upon cooling 1T-TaS$_2$ undergoes a series of different CDW transitions and at each transition the lattice goes under reconstruction and is reflected in an abrupt increase in its electrical responses. The transition from a normal undistorted metallic phase to a metallic incommensurate (IC-CDW) phase occurs at $T_{IC}$ ~ 550K, followed by nearly commensurate (NC-CDW) phase at $T_{NC}$ ~ 350K and finally, to an insulating completely commensurate state (C-CDW) which triggers in a temperature range of 180K ≤ $T_C$ ≤ 230K [23] and occurs at $T$ ~ 230K on heating [24]. C-CDW phase is well known for formation of a $\sqrt{13}$ $X$ $\sqrt{13}$ structure, known as 'Stars of David' where the sites of stars move inwards towards the middle site. It forms a triangular lattice and unit cell contains 13 Ta sites. As the valency of Ta is 4+ ($d^1$), and each Ta site has a single 5d electron. Band theory predicts metallic nature of ground state, but in contrast it is insulating which indicates towards correlation driven Mott insulator [25-27]. Recent reports have shown presence of another hidden quantum CDW state below $T_H$ ~ 80K which have revived the interest in further investigation of this material [28,29]. In particular a step like rise in the resistivity is reported between ~ 50-100K, which is much lower than the transition temperature between C-CDW and NC-CDW states [32]. Tentatively attributed to the different possible scenarios such as Mott transitions or interlayer re-ordering of the stacking structure.

The study of the dimensionality effects and interlayer coupling in different CDW phases has been an area of deep interest [30,31]. The electronic structures of bulk CDW phases strongly depend on the interlayer coupling [32-34]. The occupied states of $TaS_2$ are delocalized by the interlayer interactions and forms a quasi 1D metallic state in bulk form [35]. It has been reported that in ultrathin flakes the C-CDW phase becomes more conducting whereas NC-CDW phase becomes more metastable [36]. Further, it has been advocated that bulk and surface CDW transitions are similar for thick crystals, which means the interlayer coupling strength decreases in ultrathin limits and decouples the surface layers [32,37,38].

Clearly 1T-TaS$_2$ provides a rich theoretical playground where still lot to unveil. Raman spectroscopy has been proved to be an excellent probe to study CDWs, superconductivity and quantum spin liquids [39-42]. The presence of soft, amplitude, phase modes and discontinuities in the phonon dynamics in Raman spectra, provides a signature of CDW transitions. The quantum phenomenon becomes strong as we decrease the dimensionality of the material and consequently becomes an important tuning factor for manipulating properties of such materials. Hence, we decided to carry out an in-depth thickness dependent Raman investigation on 1T-TaS$_2$.

## 2. Results and Discussion

### 2.1 Crystal Structure details and Group Theory

1T-TaS$_2$ contains the planes of tantalum (Ta) atoms, which are surrounded by Sulphur (S) atoms in an octahedral arrangement as shown in Figure 1 (a, b). At ambient pressure and high temperatures, it possesses a triagonal symmetry [ P$\bar{3}$m1 (#164); D$_{3d}$ (-3m)]. The group theoretical prediction of the symmetry of modes is given as $\Gamma = A_g + E_g + 2E_u + 2A_{2u}$ where $\Gamma_{Raman} = A_g + E_g$ are Raman active while $\Gamma_{infrared} = E_u + A_{2u}$ are infrared active modes. Here, E-kind of modes are doubly degenerate and corresponds to in-plane atomic vibrations whereas A-kind of modes corresponds to out-of-plane vibrations [43]. Raman tensor, Wyckoff positions and corresponding phonon modes are summarized in the Table SI of supplementary information (S.I.) [44].

In the completely commensurate phase where Ta atoms form star-of-David clusters. Uchida et al., [45] reported that in this phase the new unit-cell is triclinic having $C_i^1$ symmetry and contains 39 atoms which means a total of 114 vibrational modes (57 Raman active + 57 IR-active). Polarization analysis reveals A$_g$, E$_g$ symmetry of the modes in this phase, which indicates toward a possibility of triagonal or hexagonal symmetry [46]. Further it was advocated that due to weak coupling between layers in the C-CDW phase it can even be assigned the symmetry of a single layer i.e., $C_{3i}$, which means a total of 117 modes given as : $\Gamma_{C-CDW}$ = 19 $A_g$ + 19 $E_g$ + 20 $E_u$ + 20 $A_{2u}$ [45].

**2.2 Experimental and Computational details**

We have used inelastic light scattering experiment i.e., Raman spectroscopy, using micro-Raman spectrometer (LabRAM HR Evolution) in the backscattering configuration. Sample is illuminated with a 532nm (2.33 eV) linearly polarized solid-state LASER which is focused via 50x LWD objective lens with a N.A. of 0.8. Laser power was kept low (< 0.5 mW) to prevent local heating effects. A Peltier cooled CCD detector collected the scattered light after getting dispersed by an 1800 groves/mm grating. Sample is kept under a high vacuum chamber with pressure ~ 1.0 µ torr. The sample temperature is varied over a range of 4K-330K using a closed-cycle He-flow cryostat (Montana). The temperature dependent experiment is performed on three flakes of thickness ~ 8.6 nm (T1), ~ 10.4 nm (T2) and ~ 12.5 nm (T3). These flakes are mechanically exfoliated using scotch-tape method and placed on a $SiO_2$/Si substrate. The AFM image of these flakes is shown in Figure 1(c). Further to unveil the symmetry of the phonon modes, polarization dependent study is done in a configuration where the incident light polarization direction is rotated keeping the analyzer fixed at 4K and 230K for flake T2. A pictorial representation of plane projection of this configuration is shown in Figure. 1 (d).

Structural optimization and Zone-centered phonon frequencies were calculated utilizing plane-wave approach as implemented in QUANTUM ESPRESSO package [47]. Calculations are done for 1T undistorted phase. Linear response method within Density Functional Perturbation Theory (DFPT) is used to get dynamical matrix. Ultrasoft pseudopotentials (PBESOL) are used as an exchange-correlation functional with non-linear core correction. The Kinetic-energy cutoff of 50

Ry and charge-density cutoff of 400 Ry were used. The Monkhorst-pack scheme with 15 x 15 x 15 k-point dense mesh is used for the numerical integration of the Brillouin zone. The plots and related discussion are mentioned in the S.I. [44]. We have calculated the phonon dispersion curve along the high symmetry points ($\Gamma - M - K - \Gamma - A - L - H - A | L M | H$ K) and phonon density of states (Partial and total) are also plotted for this phase as shown in Figure S1. A pictorial representation of atomic displacement of Raman active phonon modes at $\Gamma$-point is shown in Figure S2 and obtained phonon frequencies, optical activity at $\Gamma$-point is tabulated in Table SII which are also consistent with the previous calculation done by Oliver et al. [48].

## 2.3 Temperature evolution of the phonon modes

Study of temperature dependence of the lattice dynamics can provide a vital information of anharmonicity, electron-phonon coupling of respective phonon mode, structural transitions and presence of other quasi-particle excitations. We observed a total of 25 modes and a broad two phonon mode ($S_{2-ph}$) at lowest recorded temperature i.e., 4K, for all the flakes (T1, T2 and T3). Figure 2 shows comparison of raw spectra for T1, T2 and T3 at (a) 4K along with mode labels S1-S25, $S_{2-ph}$, (b) 200K and (c) 330K. Temperature evolution of raw spectra for T1, T2 and T3 flakes is shown in Figure. S3 of S.I. We have fitted the Raman spectra at different temperatures with Lorentzian function whereas $S_{2-ph}$ is fitted with a Gaussian and extracted corresponding evolution of phonon self-energy parameters (frequency, linewidth) and intensity with temperature. Interestingly, we also observed a peak at ~ 5 cm$^{-1}$ as shown in shaded orange region in Figure S3 (S.I.). It emerges and gains intensity with increase in temperature for all thickness. Its low energy suggests that it may have its origin due to interlayer interactions as sheer or breathing modes [49]. Surprisingly, it disappears with lowering the temperature, suggesting other possible origin of this ultra-low frequency mode. We suspect it may be arising due CDW effects in different phases and further detailed study is required to unveil its origin. Experimentally observed Raman active phonon modes for different thickness at 4K are mentioned in Table I.

A thickness dependent comparison of temperature evolution of phonon frequency and line-width for some modes (S5, S6, S9, S18, S21 and S23) is shown in Figure 3 and for other modes (S11, S12, S14, S15, S17 and S22) is shown in Figure S4 in S.I.. In Figure 3 we clearly observed CDW transition in the temperature evolution of the phonon modes shown by the dashed orange line for both frequency and FWHM at ~ 210K. Solid yellow and cyan line shows anharmonic and amplitude mode fit which is discussed ahead. In addition to that we observed signature of some other hidden quantum state transition at ~ 80K as clear from the Figure 3 (inset) shown in dashed red line and Figure S4. A clear change in slope is observed in the mode frequencies and linewidths for the most prominent modes S5 and S6. The possible reason for the hidden phase will be discussed in later section. The Hamiltonian for a typical CDW systems at finite temperatures can be formulated as follows $H = H_{elc.} + H_{lat.} + H_{elc.+lat.} + H_{anh.}$ here $H_{elc.} = \sum_{k} E(k)\, c^{\dagger}(k)\, c(k)$ is the electronic contribution, $H_{lat.} = \sum_{k} \hbar\omega(k)\, b^{\dagger}(k)\, b(k)$ is contribution from the lattice, $H_{elc.+lat.} = \sum_{k+q} g(q) c^{\dagger}(k+q) c(k)[b^{\dagger}(q) + b^{\dagger}(-q)]$ represents electron-lattice interaction term and $H_{anh.}$ comes into the picture due cubic and quartic anharmonic terms present in the lattice potential. $c(k)$ and $b(k)$ are electron and phonon operators; $E(k)$ and $\omega(k)$ are their respective energies; $g(q)$ is a measure of electron-phonon coupling [50]. First, we will discuss the effect of anharmonicity here. The effect of anharmonicity on frequency and FWHM that comes into picture due to presence of anharmonic terms in the lattice potential. Temperature dependence of the phonon frequencies having the contribution of three phonon i.e., one phonon decaying into two phonons of equal frequency and satisfying $\omega_1 = \omega_2 = \omega/2$; $k_1 + k_2 = 0$ and can be described using following functional form [51,52]:

$$\delta\omega_{an}(T) = \omega(T) - \omega_o = A\left(1 + \frac{2}{e^x - 1}\right) ; \qquad -- (1)$$

respectively, here $\omega_o$ is the mode frequency at absolute zero temperature; $x = \frac{\hbar\omega_o}{2k_B T}$ and $k_B$ is the Boltzmann constant. Coefficient A represents the strength of phonon-phonon interaction involving three phonon process. We observed a significant anharmonic effect in the temperature range of ~230K-330K. According to the anharmonic model generally frequency of a phonon mode increases with lowering the temperature. We fitted frequency for modes S5, S6, and S18 for T1, T2 and T3 using three-phonon model (shown for only S5 in Figure 3 as yellow solid line). The extracted parameters are mentioned in the Table SIII in S.I.

One of the important features of CDW transitions is presence of the amplitude modes which softens and broadens on warming. These CDW amplitude modes are related to the complex order parameter and follows a mean-field like temperature dependence which is given as:

$$\omega_{CDW}^i(T) = D\omega_o^i\left(1 - \frac{T}{T_{CDW}}\right)^\gamma \qquad -- (2)$$

Here $\omega_o^i$ is the unscreened, or high-temperature phonon frequency and $D \propto Ng^2(q)$ represents electron-phonon coupling constant. $N$ is joint density of states of electrons or holes involved in CDW transition, $g(q)$ is $q$ dependent electron-phonon coupling matrix element [53-55]. Solid cyan line for S9 mode in Figure 3 represent the fit and is quite good fit, where D and $\gamma$ are varied to get a reasonable value of $T_{CDW}$. We have fitted other modes which emerged below $T_{CDW}$ ~230K i.e., S2, S12, S15, S16, S20, S21 and S22. The obtained values of $\gamma$ is mentioned in the Table

SIV (S.I.). Theoretically expected value for $\gamma \sim 0.5$, but we found that this expected nature failed for the modes in our case, owing to an incomplete softening as the mode acquires a finite value of frequency instead of zero just before the phase transition as predicted. This behavior has been observed in different CDW materials and is attributed to short-range fluctuations which are out of phase with local CDW order as found for 2H-NbSe$_2$ [56] and TiSe$_2$ [55,57]. Figure 4 shows a bird's eye view for the emergence of new modes for T1 i.e., S1, S2, S3, S8, S9, S12, S13, S15, S16, S17, S19, S20, S21, S22 and S25 below ~230K. The emergence of new modes for T2 and T3 is also shown in Figure S5 (S.I.). In addition to that on further lowering the temperature we observed emergence of some weak modes i.e., S7, S10, S14 and S24 below $T_H$ ~80K. This temperature is marked by the presence of a hidden quantum CDW state.

Presence of this hidden quantum CDW state is also observed in the modes S5, S6, S18 and S23 as shown in the inset (S6 and S18) of Figure 3. Effect is also visible in the modes S11, S12, S14, S17 and S22 as shown in Figure S4 (dashed red line) (S.I.). Stojchevska et al. [29], also showed presence of this state by probing its insulating ground state using a single intense femto-second laser pulse. As compared to other states in this system the hidden state shows larger changes in the resistance, strong modification in single-particle and collective-mode spectra and is also reflected in the optical reflectivity. Stahl et al. [58], proposed that this state involves a rearrangement of the charge and orbital order in the direction of perpendicular to the TaS$_2$ layer which results in the collapse of inter-layer dimerization.

Further we observed a broad spectral feature at ~ 380 cm$^{-1}$. Such a broad Raman signatures are commonly found in CDW materials. It is associated with the strong momentum dependent electron-phonon coupling near CDW wave vector. It is a second-order process where the phonon-assisted scattering of the electrons creates two phonons with equal and opposite wave-vectors;

$\vec{q_1} + \vec{q_2} = \vec{0}$. All phonons which satisfy this condition contribute to the two-phonon scattering which leads to large FWHM. Generally multi-phonon process are weaker than one phonon process but in case of CDW materials it is found to be much stronger which is attributed to the electron-phonon coupling [59]. The temperature evolution of this mode gives an indirect measure of phonon branch renormalization during the CDW transition [60]. We label this broad two-phonon mode as $S_{2\text{-ph}}$ and the temperature evolution of its peak frequency, linewidth width and intensity are shown in Figure 5. The frequency of this mode increases linearly on decreasing temperatures till ~210K ($T_{CDW}$) and shows a discontinuity then remains nearly constant (for T3 and T2) till ~100K which is the hidden quantum phase transition temperature and finally softens till 4K for all flakes. Surprisingly, the FWHM of this mode increases with decreasing temperatures which is opposite to the normal phonon mode behavior suggesting electron-phonon coupling. The intensity variation is also anomalous as it increases with decrease in temperature.

**2.4 Polarization dependent analysis**

The Raman scattering cross-section is affected by symmetry of the crystal through second-order susceptibility/electron-phonon interaction in macroscopic/microscopic descriptions, respectively; and the intensities depends on the polarization direction of the incident light [61,62]. It can unveil the symmetry of the phonon modes, hence we performed an angle resolved polarized Raman scattering (ARPRS) experiment in a configuration where the direction of polarization of the incident light is tuned ranging from $0°$ to $360°$ using a half wave retarder ($\lambda/2$ plate), whereas analyzer principal axis has been kept fixed. The experiment is performed at two different temperatures 5K and 230K for T2 flake. Figure 6 shows angular variation in intensity and red solid line shows the fitting which is discussed below. Within semiclassical approach, the Raman scattering intensity from first-order phonon modes is related to Raman tensor, polarization configuration of incident and scattered light which can be defined as:

$$I_{Raman} = C\left[\sum_{k,l=x,y,z} e_i^k R_{kl} e_s^l\right]^2 = C\left|\hat{e}_s^\dagger \cdot R \cdot \hat{e}_i\right|^2 \quad\text{--- (3)}$$

where ' † ' symbolizes transpose, $'\hat{e}_i'$ and $'\hat{e}_s'$ are the Jones vectors representing incident and scattered light polarization direction. '$R$' represents the Raman tensor of the respective phonon mode [62,63]. A schematic diagram of polarization vectors of incident and scattered light projected on a x (*a*-axis) - y (*b*-axis) plane is shown in Figure. 1(d). In the matrix form, incident and scattered light polarization direction vectors are written as: $\hat{e}_i = [\cos(\alpha+\beta)\ \sin(\alpha+\beta)\ 0]$ ; $\hat{e}_s = [\cos(\alpha)\ \sin(\alpha)\ 0]$, where '$\beta$' is the relative angle between $'\hat{e}_i'$ and $'\hat{e}_s'$ and 'α' is an angle of scattered light from x-axis. As discussed in section 2.2 the symmetry of the modes in C-CDW and NC-CDW phase is $A_g$ / $E_g$. Hence, we have used the Raman tensors of the metallic phase with space group #164; P$\bar{3}$m1 and point group $D_{3d}$ (-3m) for fitting purpose for different modes and are summarized in Table-SI in S.I. . The angular dependency of intensities of the Raman active modes using eq. 4 can be written as $I_{A_{1g}} = |a\cos(\beta)|^2$ and $I_{E_g} = I_{E_{g,1}} + I_{E_{g,2}} = |c\cos(2\alpha+\beta)|^2 + |c\sin(2\alpha+\beta)|^2 = |c|^2$. Here $\alpha$ is an arbitrary angle from the *a*-axis and is kept constant. Therefore, without any loss of generality it can be taken as zero also.

At 5K we observed that modes S2, S3+S4, S9, S18, S19 and S21 show an $E_g$ kind of symmetry. While modes S5, S6+S7, S10, S (11-14), S15, S17, S23, S24+S25 and $S_{2-ph}$ show $A_g$ kind of symmetry. Here intensity at 90º does not goes to exact zero as theoretically predicted due experimental limitations but obtains a relative minimum value. At 230K mode S2, S3+S4, S18 and S19+S20 show similar E$_g$ kind of symmetry, while modes S5, S6+S7, S(10-14), S17, S22+S23, S24+S25 and S$_{2\text{-ph}}$ shows A$_g$ symmetry. Interestingly modes S9 and S21 showed a change in symmetry from nearly E$_g$ to A$_g$ with changing temperature. The polarization study reveals that in different CDW phases the symmetry of the modes is almost unchanged as predicted in previous reports [45]. It can be deduced that theoretically A$_g$ mode vanishes for cross configuration $I_\perp^{A_{1g}} = 0$ while E$_g$ mode intensity remains same in both parallel and crossed spectrum i.e., $I_\perp^{E_g} = I_\parallel^{E_g}$. Such an analysis within experimental limitations, can be summarized in terms of depolarization ratio $r = I_\perp / I_\parallel$ as shown in Figure S6 (S.I.) for both 5K (C-CDW) and 230K (N-CDW).

**2.5 Low frequency Raman response**

In CDW systems, gap in the electronic excitation is found to be momentum dependent [43]. The Mott transition leads to an additional opening in bands around the Γ point [27]. The opening of a gap at Fermi level reduces N$_F$, which consequently increases resistivity and this reduction of N$_F$ may manifest itself in the Raman spectra resulting in the change of initial slope of the Raman response [64,65]. Raman response; $\chi''(\omega,T)$ is calculated by dividing raw Raman intensity with Bose factor, $I(\omega,T) \propto [1+n(\omega,T)]\chi''(\omega,T)$. The Raman response, $\chi''(\omega,T)$, is the imaginary part of the susceptibility and is proportional to Stokes Raman intensity given as: $I(\omega,T) = \int_0^\infty dt\, e^{i\omega t} \langle R(t)\, R(0) \rangle \propto [1+n(\omega,T)]\chi''(\omega,T)$; where $R(t)$ is the Raman operator and $[1+n(\omega,T)] = 1/[1-e^{-\hbar\omega_b/k_B T}]$ is the Bose factor. The initial slope of the Raman response is $R \lim_{\omega \to 0} \frac{\partial \chi''}{\partial \omega} \propto N_F \tau_o \propto \frac{1}{\Gamma_0^*}$, here R incorporates experimental factors and $\Gamma_0^*$ represent electronic scattering rate. The scattering rate of the conduction electrons also referred as Raman resistivity [66]. It is anticipated that with increasing correlation at low temperature, as the system under study goes from metal to a correlated one and finally to a Mott insulating state inside the CDW phase, lifetime of putative quasiparticles decreases and as a result inverse of slope ($\Gamma_0^*$) should increase. Figure 7 shows the inverse of slope for all thickness as a function of temperature. To extract the slope, we linearly fitted the low frequency region in the range 8 cm$^{-1}$ – 21 cm$^{-1}$. We observe that with decreasing temperature it increases till ~ 210K and then start decreasing till ~ 100K and with further lowering the temperature it increases below ~ 50-60K. With decreasing temperature below T$_{CDW}$, there is a small but finite *U* (onsite correlations) contribution and the correlated metal displays an inverse slope $\propto T^2$ reflecting a canonical Fermi liquid behaviour i.e. correlated metallic phase. Interestingly, on further decreasing the temperature below ~ 50-60K we saw an upturn in the inverse of slope suggesting that system becomes more insulating with decreasing the temperature.

### 2.5.1 Quantum spin liquids

Quantum Spin Liquid (QSL) is a state in which the quantum fluctuations of the spins are strong enough to preclude the ordering even at absolute zero. Such a state was first envisioned by P.W. Anderson in 1973 and later inspired researchers to study the frustrated magnetic systems. The key feature of spin liquids is the presence of fractionalized excitations. Such quantum particles carry spin ½ but lacks charge and are known as spinons [67]. 1T-TaS$_2$ in its C-CDW state forms star of David structure with 12 Ta-atoms pair and forms 6 occupied bands, which leaves central Ta atom localized and unpaired. The band structure calculations in the commensurate state have pointed that mostly Ta 5d orbitals contribute in the conduction and valence bands [31,68], while the atomic spin-orbit coupling (SOC) from $d_{x^2-y^2}$ and $d_{xy}$ orbitals go under reconstruction and modify the band structure. This combined effect of structural deformation and atomic SOC give rise to a narrow band gap at the Fermi level. Thus, there is a weak repulsive interaction and the resulting ground state has been proposed to be a Mott insulator [68-70]. Mott insulators generally shows an antiferromagnetic ordering in the insulating phase which is not in the case of 1T-TaS$_2$. It has been shown that due to large paramagnetic contribution to the magnetic susceptibility it does not show any magnetic ordering even down to 20mK [71]. The localized unpaired electrons carry spin-1/2 and form a triangular lattice which makes it a putative candidate for the realization of quantum spin liquid [70-72]. The experimental realization of QSL has been an active area of research and Raman spectroscopy has proved to be an excellent probe for that [41,73]. Magnetic Raman scattering give rise to a broad continuum originates from underlying dynamical spin fluctuations and may be used to gauge the fractionalization of quantum spins expected for proximate spin liquid candidates [41,73-79].

Hence, we have analyzed the possible existence of very much anticipated QSL state and effect of dimensionality for 1T-TaS$_2$ via Raman Spectroscopy. Signatures of a QSL can be uncovered in the presence of underlying dynamic quantum spin and/or orbital fluctuations and Raman technique is an excellent probe for probing these underlying dynamical fluctuations and is reflected in the emergence of quasi-elastic peak at low frequency and $\chi^{dyn}(\omega,T)$ [41,77,79]. We observed a profound temperature variation in the low frequency region which we consider the quasi-elastic part. It arises from diffusive fluctuations of a four-spin time correlation function or fluctuations of the magnetic energy density [80]. The temperature dependence of quasi-elastic scattering intensity can provide information about the evolution of low-energy excitations. The analysis of QES is done by analyzing the temperature dependence of intensity under the curve for the low frequency range of ~ 6 cm$^{-1}$ - 21 cm$^{-1}$. Figure 8 (a) shows the temperature dependent raw spectra and Raman conductivity for T1. The variation of raw intensity spectral weight with temperature for different thickness is shown in Figure 8 (b). In addition to that we also observed signature of the proposed hidden phase as discontinuity at T$_H$ ~ 80K, which has also been reflected in the phonon dynamics. We found another possible signature of a quantum transition below 50K as shown in shaded violet region, which may be related to the quantum spin liquid and it needs further investigation.

For further probing the evolution of the underlying quantum spin fluctuations we quantitatively evaluated $\chi^{dyn}$, shown in Figure. 8 (c). $\chi^{dyn}$ is evaluated using Kramers - Kronig relation which relates the real and imaginary part of the susceptibility as:

$$\chi^{dyn} = \lim_{\omega \to 0} \chi(k=0,\omega) \equiv \frac{2}{\pi} \int_0^\Omega \frac{\chi''(\omega)}{\omega} d\omega \tag{5}$$

Where $\chi''(\omega)/\omega$ is Raman conductivity, $\Omega$ is the upper cutoff value of integrated frequency and is taken as 21 cm$^{-1}$ for the Quasi-elastic scattering (QES) region. For all thickness with lowering temperatures other than small discontinuities at T$_{CDW}$ ~ 200K which is CDW transition temperature $\chi^{dyn}$ remains almost independent behavior which is expected in a paramagnetic regime where spins remain uncorrelated. Interestingly it starts to increase drastically on further lowering the temperature below, ~ 50K for T1, T2 and T3. We consider these temperatures as the onset point of dynamical fluctuation as cross-over temperature. For a typical antiferromagnets $\chi^{dyn}$ is expected to quench below Néel temperature (T$_N$). Such a behaviour is a typical characteristic scattering from underlying quantum spin fluctuations [41,78]. The variation of $\chi^{dyn}$ is well described by power law $\chi^{dyn} \propto T^\alpha$. The power law is associated with the inherent long-range entangled spin-liquid and is very different from conventional magnets. Hence, we have fitted it with power law as shown in the Figure 9 (c) and the obtained values of $\alpha$ for T1, T2 and T3 are -0.65 $\pm$ 0.07; -0.56 $\pm$ 0.16 and -0.51 $\pm$ 0.14, respectively. Here $\alpha$ reflects on the underlying valence bond randomness due to spins [81] and it increases with decreasing thickness which suggests increasing quantum fluctuations in low dimensions. Further, we inquired the polarization dependence of the quasi-elastic $\chi^{dyn}$ response for T2 flake at 5K and 230K which is shown in Figure 8 (d). At both temperatures it shows an $A_g$ kind of symmetry i.e., two-fold rotational anisotropy similar to the $A_g$ phonon modes.

**Conclusion**

In conclusion, we performed a detailed temperature (4K-330K) and polarization dependent Raman measurements on three flakes with varying thickness of 1T-TaS$_2$. A clear signature of CDW transition is seen as we observed the emergence of new modes on going from the high temperature NC-CDW towards the low temperature C-CDW phase. We observed discontinuities in the phonon dynamics around T$_{CDW}$. In addition to the CDW transition, strong signature of a hidden quantum CDW state is also observed at $T_H$ ~ 80K marked by the emergence of new modes and discontinuities in the most prominent modes. Our polarization dependent analysis shows that the symmetry of most of the modes remains same even in the CDW phases which is consistent with previous reports. Our analysis of the low frequency quasi-elastic region also shows that there are some additional competing interactions at low temperatures, which is reflected in the slope of Raman response. The analysis of dynamic fluctuations qualitatively showed by $\chi^{dyn}$ and the power law fit reflects increase in quantum fluctuation effects on decreasing thickness. Our results show a possible Mott insulating state below ~ 50-60 K. Our results reflect on the interesting interplay of dimensionality with the underlying hidden quantum state as well as normal CDW states in 1T-TaS$_2$.

**Acknowledgments**

PK acknowledge support from IIT Mandi for the experimental facilities and financial support from SERB (CRG/2023/002069) India.

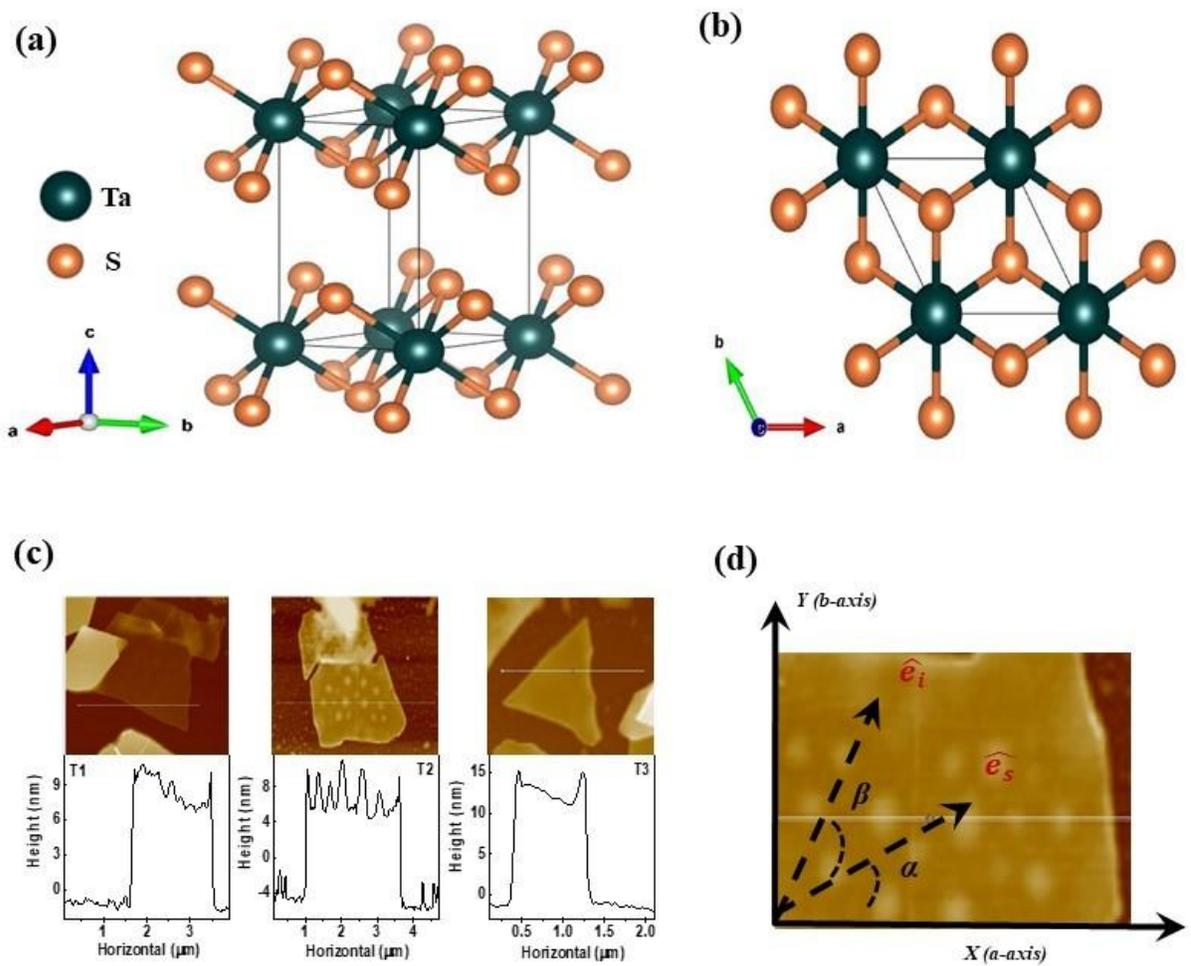

**Figure 1:** Shows structure of the undistorted phase of 1T-TaS$_2$ in **(a)** standard orientation **(b)** along c-axis **(c)** AFM Image of T1, T2 and T3 and **(d)** Plane projection of polarization direction of incident and scattered light.

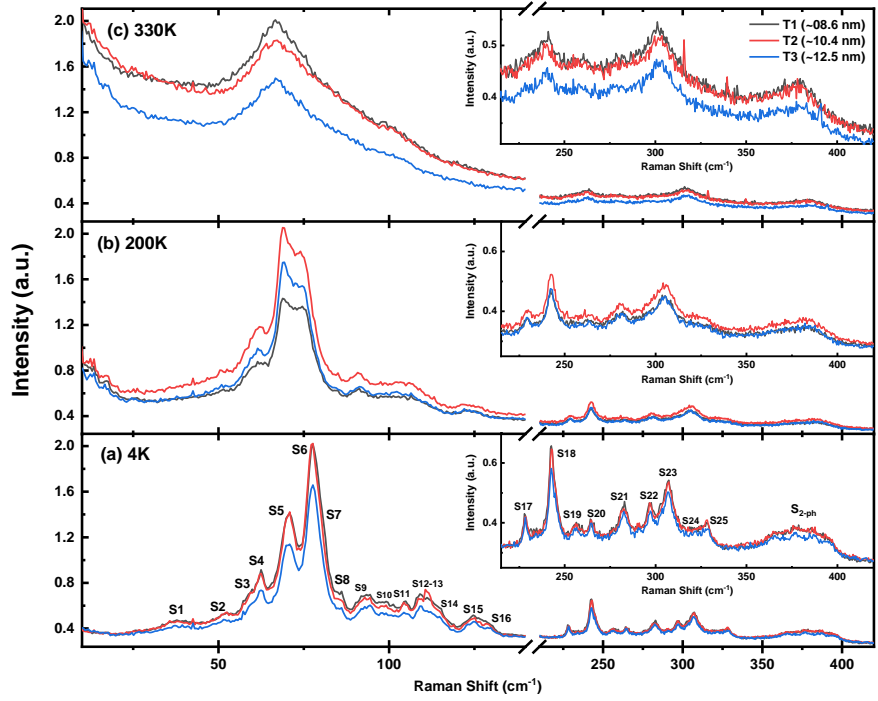

**Figure 2:** Raman Spectra for T1, T2 and T3 at 4K, 200K and 330K along with mode labels.

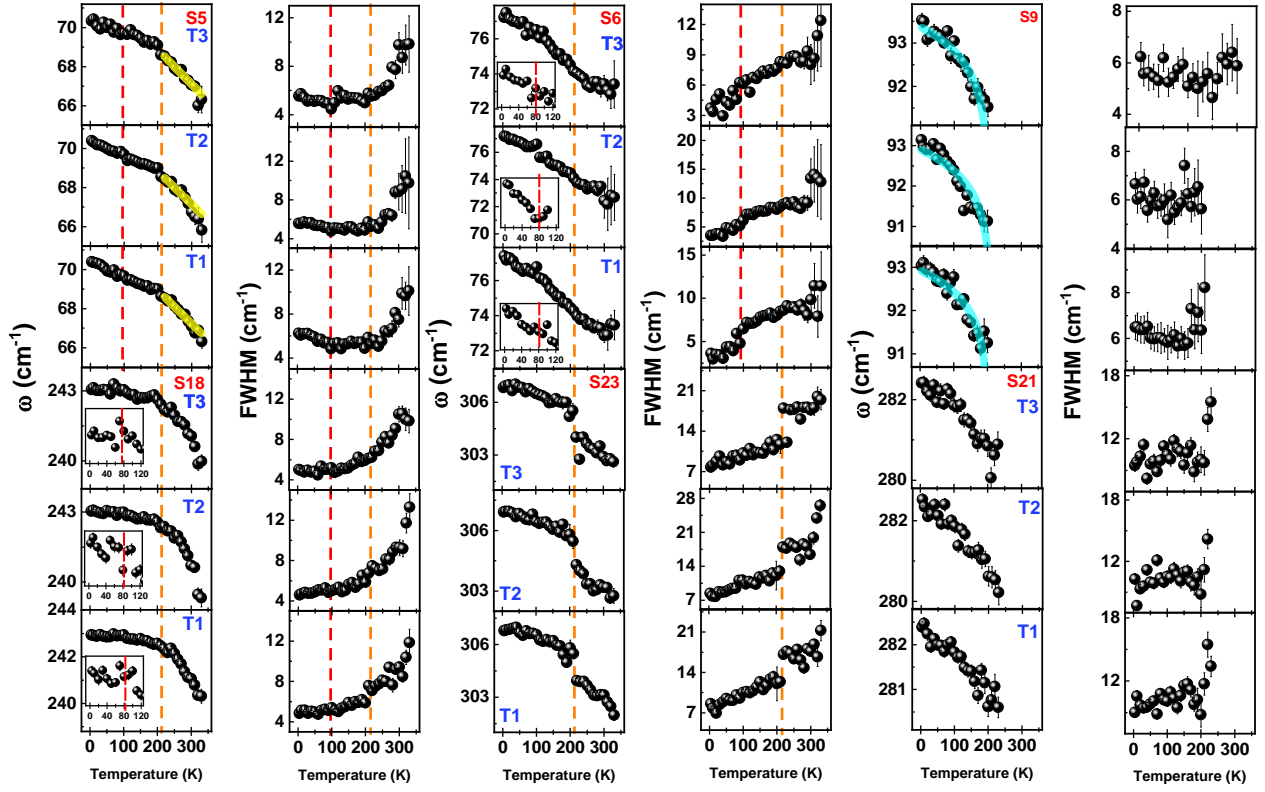

**Figure 3.** Temperature dependent frequency and linewidth for T1, T2 and T3 of some of the modes i.e., S5, S6, S9, S18, S21 and S23. Yellow and cyan solid line shows fit as mentioned in the text. Red and orange dashed line shows $T_H$ and $T_{CDW}$ transition temperature.

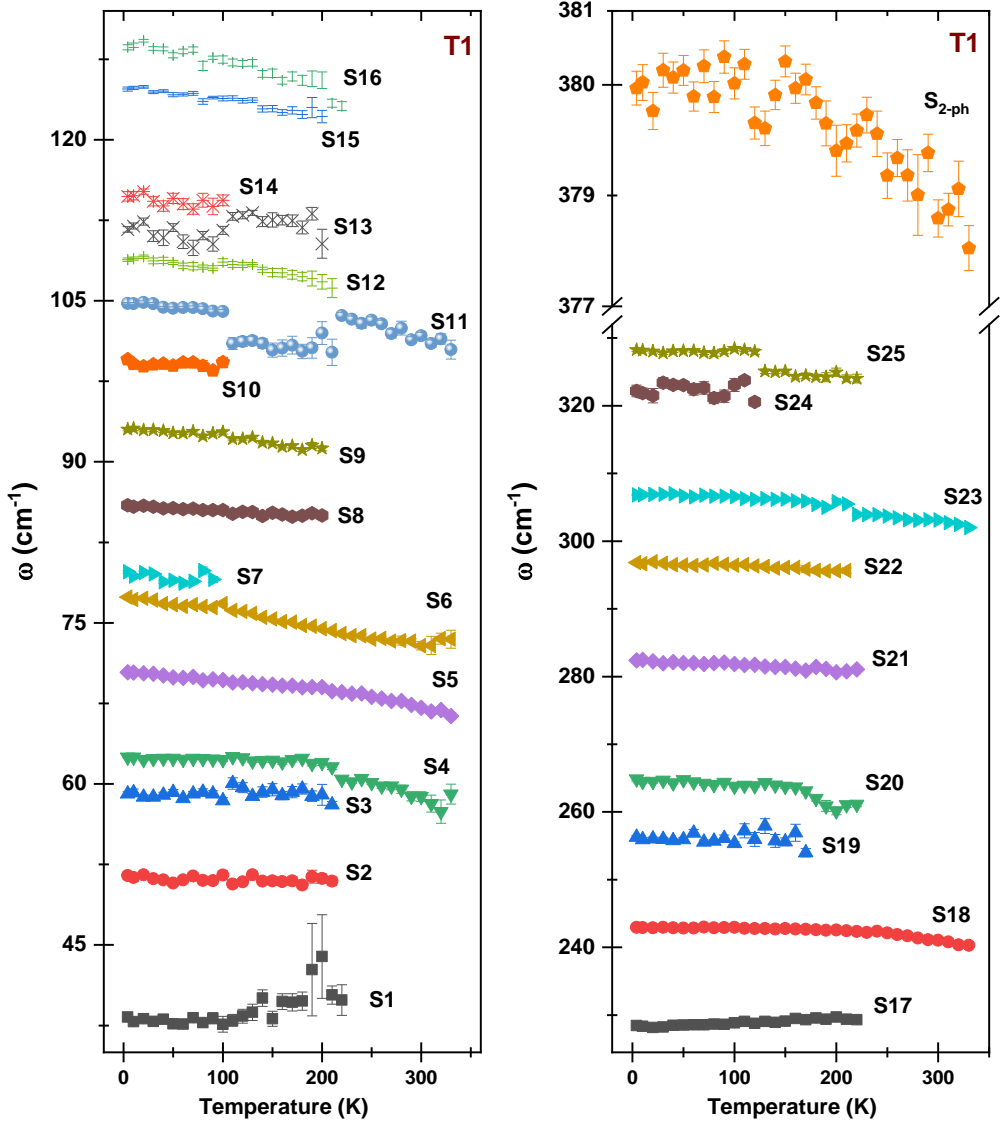

**Figure 4:** Temperature evolution of the observed phonon modes frequency for T1.

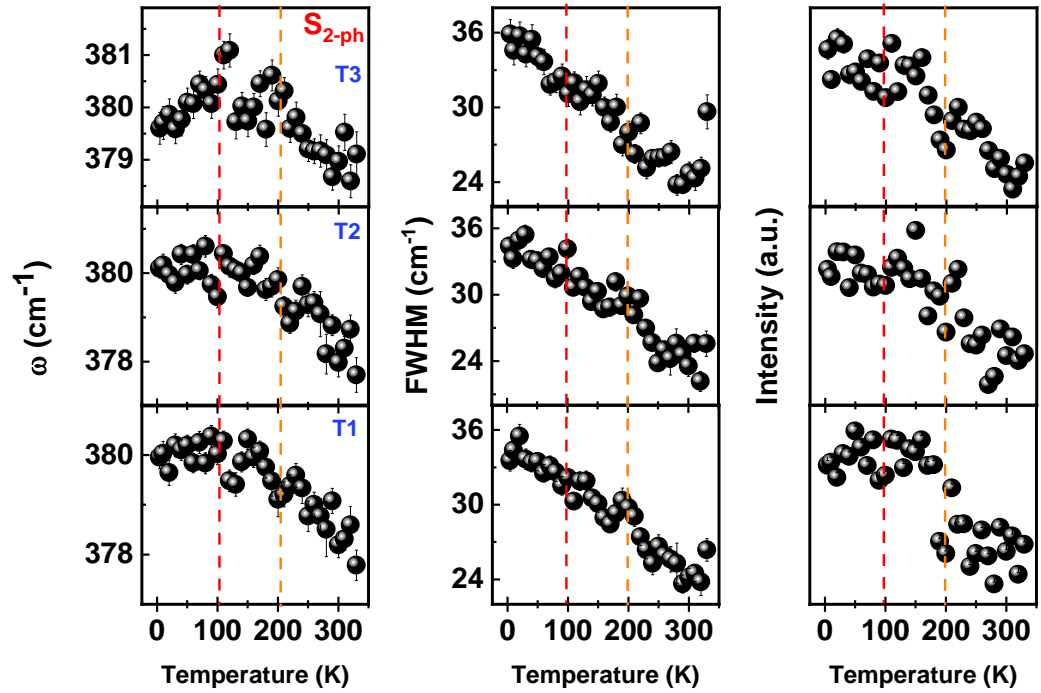

**Figure 5:** Temperature dependent frequency and linewidth variation for T1, T2 and T3 of phonon modes $S_{2\text{-ph}}$. Red and orange dashed line shows $T_H$ and $T_{CDW}$ transition temperature.

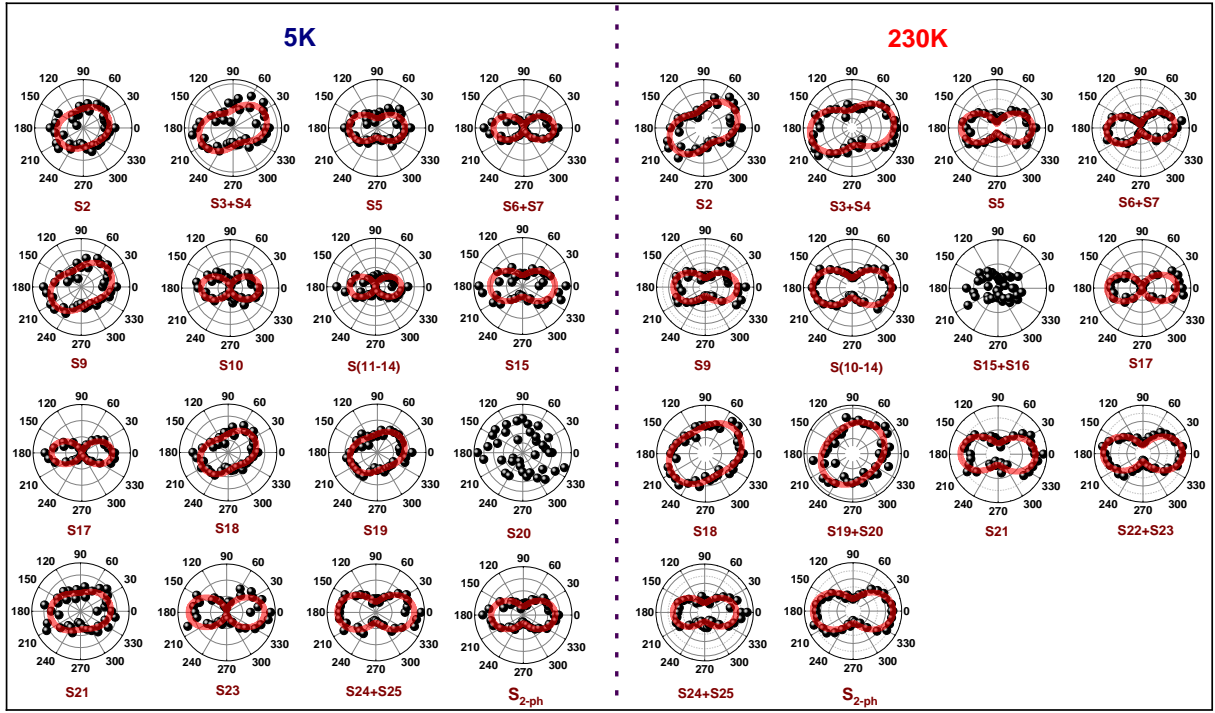

**Figure 6**: Polarization dependent intensity variation of modes at 5K and 230K for T2. Red solid line is a fit as mentioned in the text.

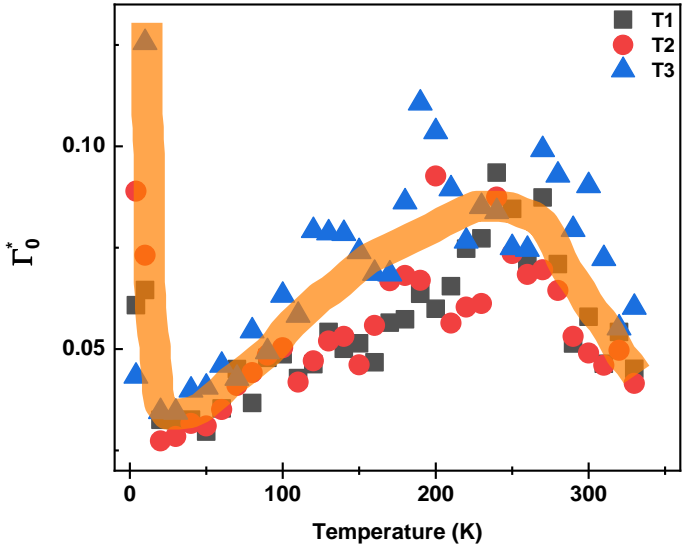

**Figure 7**: Raman response slope magnitude variation with temperature for T1, T2 and T3. Orange solid line is guide to the eye.

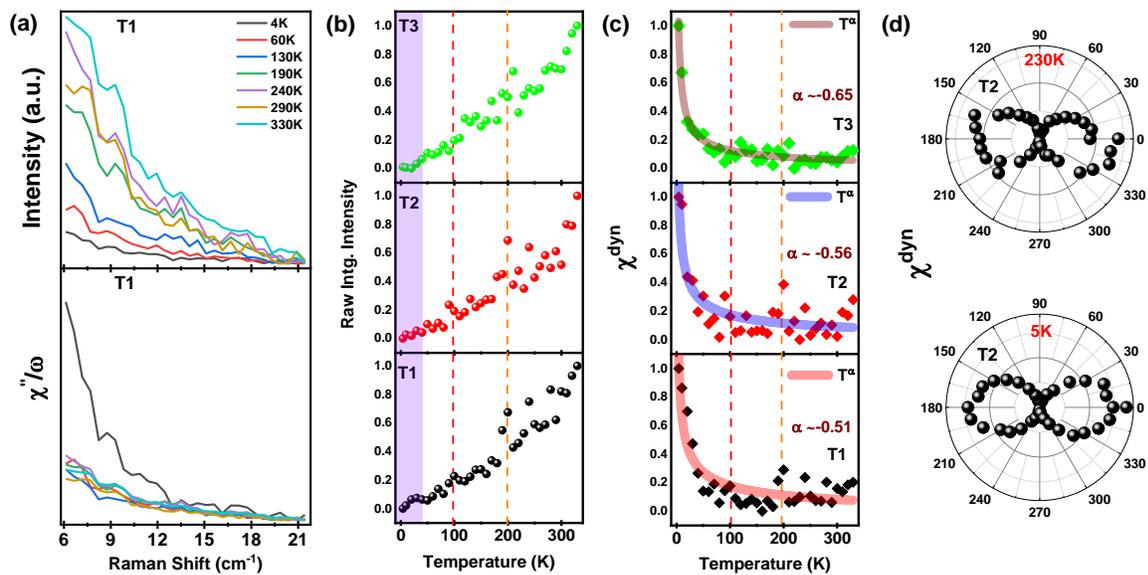

**Figure 8:** **(a)** Raw spectra and Raman conductivity of Low frequency region; **(b)** Integrated Raw intensity and **(c)** $\chi^{dyn}$. Thick solid lines are the power law fit as mentioned in the text and **(d)** polarization dependent Raw spectra, Raman conductivity and $\chi^{dyn}$ at 5K and 230K for Low frequency region for flake T2.

**Table I:** Experimentally observed modes frequency ($\omega_{EXP}$) in cm$^{-1}$ at 4K for different thickness i.e., T1, T2 and T3 and their tentative symmetry assignment.

| Modes # | $\omega_{EXP.}$ T1 | $\omega_{EXP.}$ T2 | $\omega_{EXP.}$ T3 | Modes # | $\omega_{EXP.}$ T1 | $\omega_{EXP.}$ T2 | $\omega_{EXP.}$ T3 |
|---|---|---|---|---|---|---|---|
| S1 | 38.3 ± 0.3 | 38.0 ± 0.3 | 38.3 ± 0.3 | S15 ($A_g$) | 124.7 ± 0.2 | 124.6 ± 0.1 | 124.8 ± 0.1 |
| S2 ($E_g$) | 51.5 ± 0.3 | 51.7 ± 0.3 | 51.3 ± 0.3 | S16 | 128.6 ± 0.2 | 128.9 ± 0.2 | 129.4 ± 0.2 |
| S3 ($E_g$) | 59.1 ± 0.2 | 58.88 ± 0.2 | 59.5 ± 0.3 | S17 ($A_g$) | 228.4 ± 0.1 | 228.6 ± 0.1 | 228.4 ± 0.1 |
| S4 ($E_g$) | 62.5 ± 0.1 | 62.3 ± 0.1 | 62.5 ± 0.1 | S18 ($E_g$) | 242.9 ± 0.0 | 243.0 ± 0.0 | 243.1 ± 0.0 |
| S5 ($A_g$) | 70.4 ± 0.0 | 70.4 ± 0.0 | 70.3 ± 0.0 | S19 ($E_g$) | 256.3 ± 0.3 | 256.2 ± 0.3 | 256.2 ± 0.3 |
| S6 ($A_g$) | 77.4 ± 0.0 | 77.2 ± 0.0 | 77.2 ± 0.1 | S20 | 264.9 ± 0.3 | 264.9 ± 0.2 | 264.8 ± 0.2 |
| S7 ($A_g$) | 79.8 ± 0.1 | 79.4 ± 0.1 | 78.9 ± 0.4 | S21 ($E_g$) | 282.4 ± 0.1 | 282.5 ± 0.1 | 282.4 ± 0.1 |
| S8 | 85.9 ± 0.1 | 85.9 ± 0.1 | 85.9 ± 0.1 | S22 | 296.8 ± 0.1 | 296.9 ± 0.1 | 296.9 ± 0.2 |
| S9 ($E_g$) | 93.0 ± 0.1 | 93.1 ± 0.1 | 93.5 ± 0.2 | S23 ($A_g$) | 306.8 ± 0.1 | 306.9 ± 0.1 | 306.8 ± 0.1 |
| S10 ($A_g$) | 99.5 ± 0.2 | 99.2 ± 0.2 | 99.6 ± 0.4 | S24 ($A_g$) | 322.2 ± 0.9 | 322.9 ± 0.8 | 322.8 ± 0.9 |
| S11 ($A_g$) | 104.7 ± 0.1 | 104.6 ± 0.1 | 104.7 ± 0.2 | S25 ($A_g$) | 328.3 ± 0.2 | 328.2 ± 0.1 | 328.5 ± 0.2 |
| S12 ($A_g$) | 108.8 ± 0.2 | 108.0 ± 0.1 | 109.1 ± 0.1 | $S_{2-ph}$($A_g$) | 379.9 ± 0.2 | 380.1 ± 0.2 | 379.6 ± 0.3 |
| S13 ($A_g$) | 11.6 ± 0.3 | 111.2 ± 0.1 | 112.05 ± 0.3 | | | | |
| S14 ($A_g$) | 114.7 ± 0.4 | 114.7 ± 0.2 | 115.1 ± 0.3 | | | | |

**Supplementary Information:**

# Hidden Quantum State and Signature of Mott Transition in Two-dimensional 1T-TaS$_2$

Vivek Kumar[*,1], Birender Singh[*,2] and Pradeep Kumar[*,3]

[*]*School of Physical Sciences, Indian Institute of Technology Mandi, Mandi-175005, India*

[1]vivekvke@gmail.com
[3]pkumar@iitmandi.ac.in
[2]Current affiliation: Department of Physics, Boston College, Chestnut Hill, MA, USA.

## S1: DFT calculations

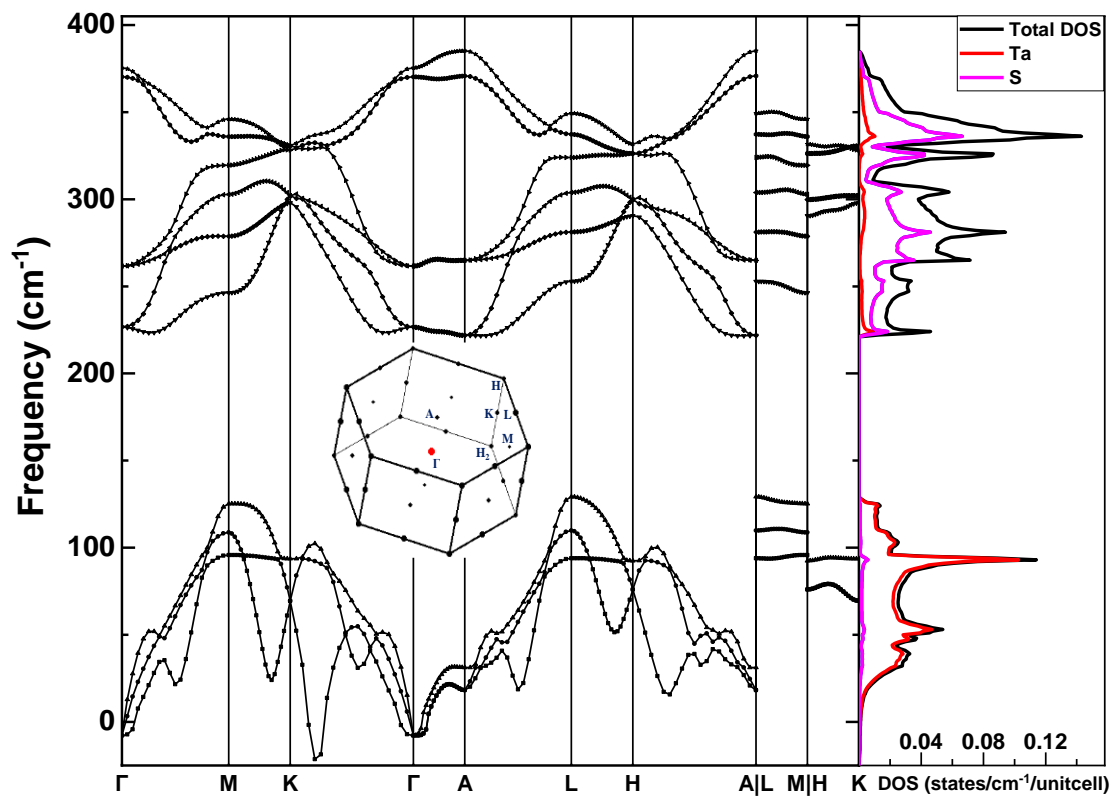

**Figure S1:** Phonon dispersion and DOS for 1T undistorted phase.

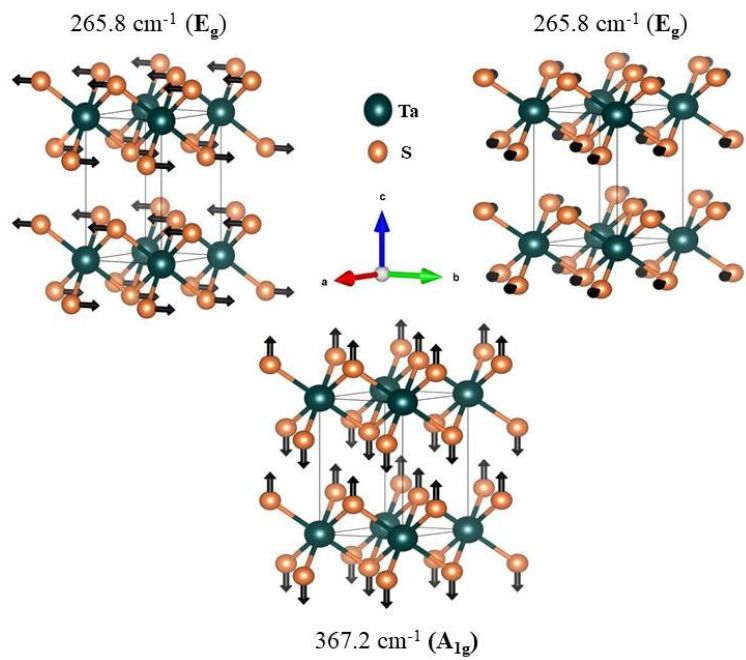

**Figure S2:** Pictorial representation of Raman active modes atomic displacement at the $\Gamma$-point in 1T undistorted phase.

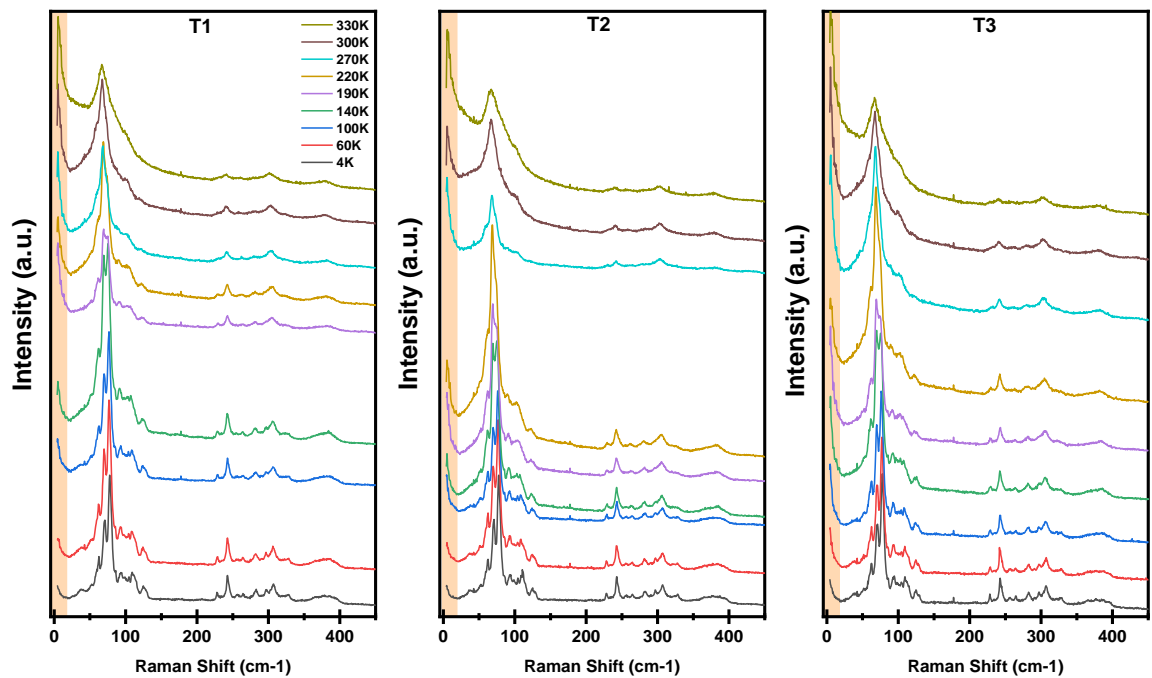

**Figure S3:** Temperature evolution of the Raman Spectra for T1, T2 and T3. Shaded orange area shows low frequency peak.

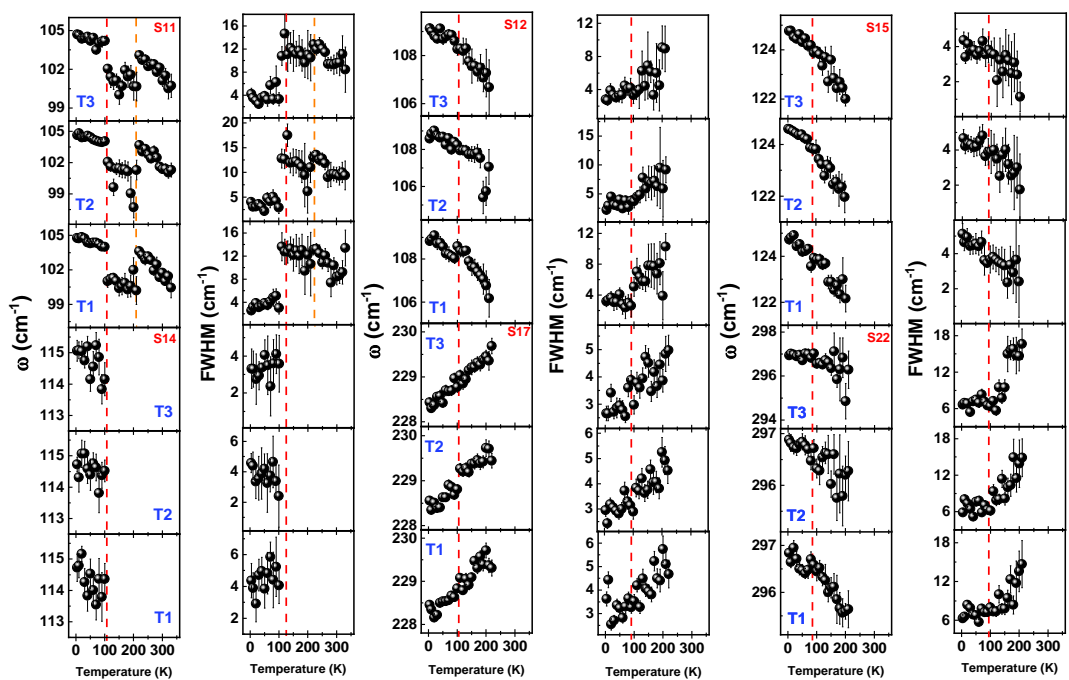

**Figure S4:** Temperature dependent frequency and linewidth variation for T1, T2 and T3 of some of the modes S11, S12, S14, S15, S17 and S22. Red and orange dashed line shows $T_H$ and $T_{CDW}$ transition.

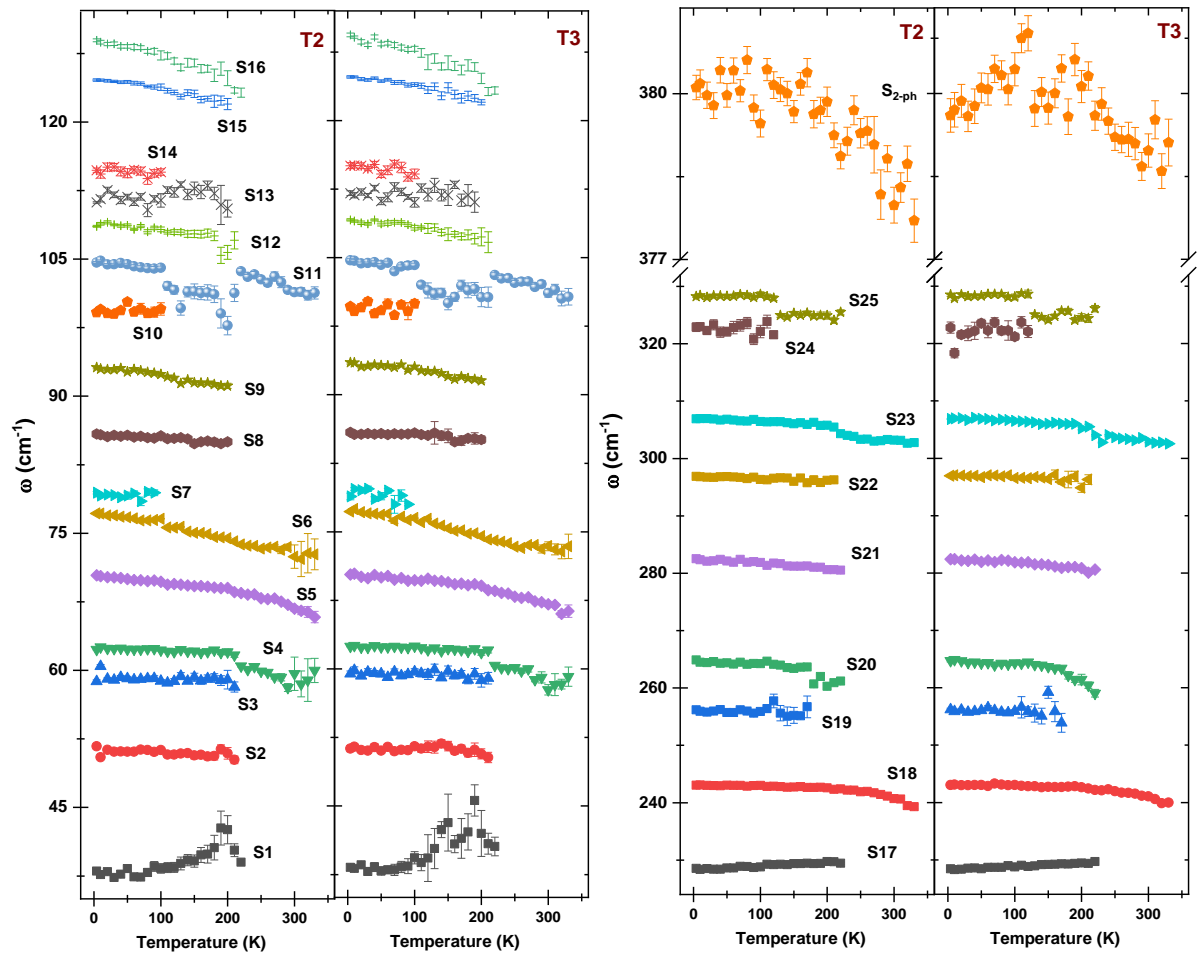

**Figure S5:** Temperature variation of frequency for T2 and T3 of all modes.

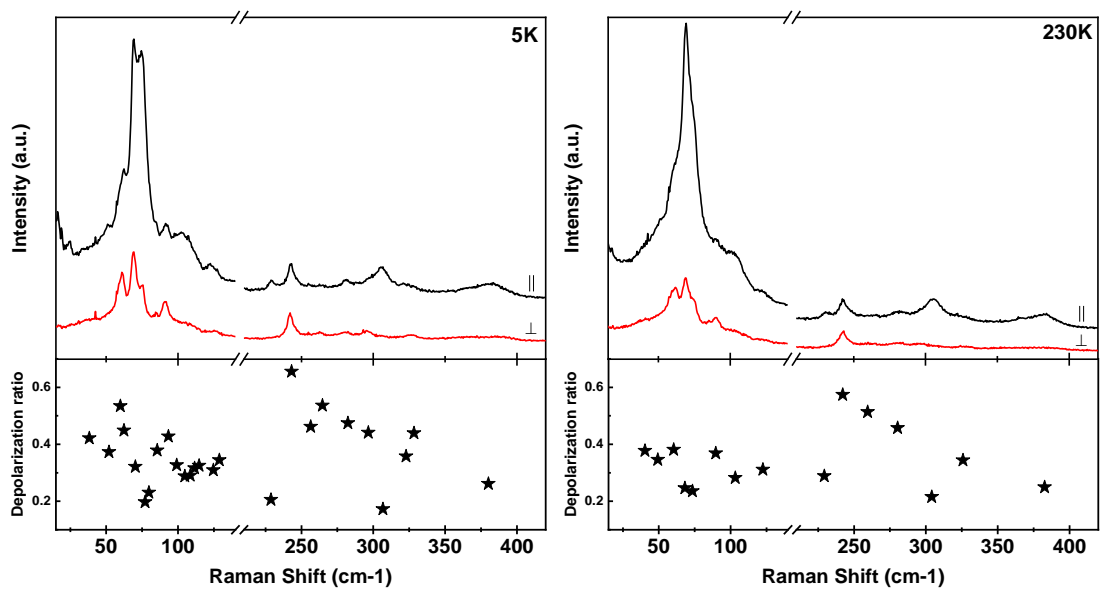

**Figure S6**: Depolarization ratio of intensity of raw spectra at 5K and 230K for T2. Modes with higher depolarization ratio suggest $E_g$ symmetry and those with lower ratio as $A_g$ symmetry.

**Table SI:** Wyckoff positions, irreducible representations of the phonon modes of trigonal [#164; P$\bar{3}$m1; D$_{3d}$ (-3m)] 1T-TaS$_2$ at the gamma point and Raman Tensors of Raman active phonon modes.

| Atoms | Wyckoff site | Γ-point mode decomposition | Raman Tensors |
|---|---|---|---|
| Ta | 1a | $A_{2u}\,(I.R) + E_u\,(I.R)$ | $A_g = \begin{pmatrix} a & 0 & 0 \\ 0 & a & 0 \\ 0 & 0 & b \end{pmatrix}$ |
| S | 2d | $A_{2u}\,(I.R) + E_u\,(I.R) + A_g(R) + E_g(R)$ | $E_{g,1} = \begin{pmatrix} c & 0 & 0 \\ 0 & -c & d \\ 0 & d & 0 \end{pmatrix}; E_{g,2} = \begin{pmatrix} 0 & -c & -d \\ -c & 0 & 0 \\ -d & 0 & 0 \end{pmatrix}$ |
| | $\Gamma = A_g + 2A_{2u} + 2E_u + E_g$ | $\Gamma_{\text{Raman}} = A_g + E_g$ | $\Gamma_{\text{Infrared}} = 2A_{2u} + 2E_u$ |

**Table SII:** DFT calculation for $D_{3d}$ point group (Bulk).

| Modes # (Symmetry) | $\omega_{DFT}$ (cm$^{-1}$) | Optical Activity |
|---|---|---|
| 1 ($A_{2u}$) | -7.8 | I |
| 2 ($E_u$) | -7.6 | I |
| 3 ($E_u$) | -7.6 | I |
| 4 ($E_u$) | 229.4 | I |
| 5 ($E_u$) | 229.4 | I |
| 6 ($E_g$) | 265.8 | R |
| 7 ($E_g$) | 265.8 | R |
| 8 ($A_{1g}$) | 367.2 | R |
| 9 ($A_{2u}$) | 375.5 | I |

**Table SIII:** Value of anharmonic fit parameters for different thickness as mentioned in the text.

| Modes | T1 | | T2 | | T3 | |
|---|---|---|---|---|---|---|
| # | $\omega_o$ | A | $\omega_o$ | A | $\omega_o$ | A |
| S5 | 72.4 ± 0.3 | -0.4 ± 0.0 | 72.3 ± 0.4 | -0.4 ± 0.1 | 72.5 ± 0.3 | -0.5 ± 0.0 |
| S6 | 76.3 ± 0.3 | -0.3 ± 0.0 | 75.5 ± 0.4 | -0.2 ± 0.0 | 76.6 ± 0.5 | -0.3 ± 0.1 |
| S18 | 246.8 ± 0.5 | -1.6 ± 0.2 | 247.2 ± 0.6 | -1.8 ± 0.2 | 246.3 ± 0.6 | -1.5 ± 0.2 |

**Table SIV:** Value of $\gamma$ for modes corresponding to different thickness as mentioned in the text.

| Modes # | $\gamma$ $10^{-3}$ T1 | $\gamma$ $10^{-3}$ T2 | $\gamma$ $10^{-3}$ T3 |
|---|---|---|---|
| S9  | 7.8 ± 2.1 | 8.8 ± 2.2 | 8.7 ± 1.8 |
| S12 | 8.1 ± 1.2 | 6.9 ± 1.4 | 9.6 ± 1.2 |
| S15 | 2.3 ± 1.5 | 8.5 ± 1.2 | 5.8 ± 1.4 |
| S16 | 9.9 ± 2.7 | 9.3 ± 2.0 | 9.3 ± 2.6 |
| S20 | 4.3 ± 0.8 | 3.8 ± 1.3 | 3.7 ± 0.6 |
| S21 | 2.5 ± 0.7 | 2.8 ± 0.6 | 2.6 ± 0.7 |
| S22 | 1.3 ± 0.2 | 1.0 ± 0.3 | 0.9 ± 0.4 |